\documentclass[aps,prl,twocolumn,showpacs,floatfix,superscriptaddress]{revtex4}
\usepackage{dcolumn}
\usepackage{amsmath}
\usepackage{amsfonts}
\usepackage{amssymb}
\usepackage{graphicx}
\usepackage{times}
\begin{document}

\title{Three-dimensional MgB$_{2}$-type superconductivity in hole-doped diamond.}
\author{Lilia Boeri}
\affiliation{INFM Center for Statistical Mechanics and Complexity and Dipartimento di
Fisica, Universit\`a di Roma ``La Sapienza'', Piazzale A. Moro 2, 00185 Roma, Italy}
\affiliation{Max-Planck Institut f\"{u}r Festk\"{o}rperforschung, Heisenbergstr. 1, D-70569
Stuttgart, Germany}
\author{Jens Kortus}\email{j.kortus@fkf.mpg.de}
\author{O. K. Andersen}
\affiliation{Max-Planck Institut f\"{u}r Festk\"{o}rperforschung, Heisenbergstr. 1, D-70569
Stuttgart, Germany}
\date{\today}

\begin{abstract}
We substantiate by numerical and analytical calculations that the recently 
discovered superconductivity
below 4 K in 3\% boron-doped diamond is caused by electron-phonon coupling of
the same type as in MgB$_{2},$ albeit in 3 dimensions. Holes at the top of the
zone-centered, degenerate $\sigma$-bonding valence band couple strongly to the
optical bond-stretching modes. The increase from 2 to 3 dimensions reduces the
mode-softening crucial for $T_{c}$ reaching 40 K in MgB$_{2}.$ Even if diamond
had the same \emph{bare} coupling constant as MgB$_{2},$ which could be
achieved with 10\% doping, $T_{c}$ would only be 25 K. Superconductivity above
1 K in Si (Ge) requires hole-doping beyond 5\% (10\%).
\end{abstract}
\maketitle

Recently, superconductivity below $T_{c}\sim$4 K was reported in diamond doped
with $x\sim$3\% boron, that is, with $\sim$0.03 holes per carbon atom
\cite{Ekimov}. Such high hole-doping levels can be achieved due to the small
size of boron. It had previously been observed that
the prominent Raman line 
caused by the zone-center optical phonons at 1332 cm$^{-1}$
downshifts and broadens significantly upon heavy boron doping \cite{Zhang}. In
this Letter we shall make plausible that the superconductivity in hole-doped
diamond is due to the coupling of the holes to the optical zone-center
phonons, a mechanism similar to the one causing high-temperature
superconductivity in MgB$_{2},$ but without some of its interesting features.
We shall also estimate transition temperatures for hole-doped Si and Ge.

The discovery \cite{Akimitsu} of superconductivity below 40 K in MgB$_{2},$ a
binary compound isostructural and isoelectronic with graphite, came as a
surprise for the scientific community. By now, it is well understood what the
mechanism is and why MgB$_{2}$ is special
\cite{Kortus,Pickett,Kong,Liu,Canfield}: In contrast to other known $sp^{2}
$-bonded superconductors, such as intercalated graphite, alkali doped
fullerides, and organic superconductors whose charge carriers are exclusively
$\pi$-electrons, MgB$_{2}$ has holes at the top of the bonding $\sigma$-bands
at the zone center. These holes, on two narrow Fermi cylinders with radii
$\sim$1/5 of the Brillouin-zone radius, couple strongly $\left(  \lambda
\sim1\right)  $ to the two optical bond-stretching modes with $q\leq2k_{F}\ll
k_{BZ},$ giving rise to a strong 2-dimensional Kohn anomaly in the phonon
spectrum. This strong coupling between a few zone-center holes and optical
phonons is what drives the high-temperature superconductivity in MgB$_{2}$.
Experience shows \cite{Pickett}, and it can be proved 
for parabolic bands with $2k_{F}\ll k_{BZ}$ \cite{Andersen}, 
that the coupling constant is given by the Hopfield expression,
\begin{equation}
\lambda=\frac{ND^{2}}{M\omega^{2}},\label{e1}
\end{equation}
where $N$ is the density of states (DOS) per spin at the Fermi level of the
$\sigma$ holes. Moreover, $\pm Du$ is the splitting of the degenerate top of
the $\sigma$-band by the displacement $\mathbf{e}u$ of a frozen, optical
zone-center phonon with normalized eigenvector $\mathbf{e}$ and energy
$\omega$. The optical phonons are softened by the interaction with the holes,
$\omega^{2}\sim\omega_{0}^{2}/\left(  1+2\lambda\right)  $ when $q<2k_{F},$
and that significantly enhances $\lambda$ and $T_{c}\sim\omega
\exp\left(  -1/\lambda\right)  $. 
This softening is presumably weakened by anharmonicity \cite{Choi,Mauri}. 
The DOS is independent of doping
because the $\sigma$-band is 2-dimensional. As a consequence, a decrease in
the number of holes, e.g. by carbon-doping, should not cause $\lambda$ to
decrease, except through the anharmonic hardening of $\omega$ caused by the
decrease of $E_{F}$ \cite{Boeri}.
In stoichiometric MgB$_{2}$ there are more carriers in the $\pi$ bands
than in the  $\sigma$ bands (0.09 per B), but the former couple far less to
phonons than the latter, and since there seems to be very little impurity
scattering between the $\sigma$- and $\pi$-bands, MgB$_{2}$ is the first
superconductor which clearly exhibits multiple gaps below a common $T_{c}$
\cite{Liu,Canfield,Mazin,Choi}.

Instead of having $\pi$-bands and three 2-dimensional bonding $\sigma$-bands,
$sp^{3}$-bonded semiconductors like diamond have four 3-dimensional 
bonding $\sigma$-bands.
The top of this valence band is three-fold degenerate
with symmetry $T_{2g},$ and so are the zone-center optical phonon modes. 
Like in MgB$_{2},$ $\sigma$-holes with
small $k_{F}$ should therefore couple strongly, and for small $k_{F}$
exclusively, to the optical bond-stretching modes, with the main differences
being that in 3 dimensions the Kohn anomaly is weaker and the DOS increases with
hole doping like $k_{F},$   the radius of the average Fermi
sphere. Since there are 3 bands and two carbon atoms per cell, $\left(
k_{F}/k_{BZ}\right)  ^{3}=x/3.$ For $x$=0.03, $k_{F}/k_{BZ}$ is 0.22, which
is like in MgB$_{2}.$ Due to the lack of a metallic $\pi$-band, diamond
becomes an insulator  once $x$ is below 1-2 per cent. We shall now
substantiate this scenario for the observed superconductivity in hole-doped
diamond by providing quantitative details, and we shall also 
consider the possibility of superconductivity in hole-doped Si and Ge. 
In particular, we shall present results of
density-functional (LDA) calculations and estimate $T_{c}$ using Eliashberg theory.

A substitutional boron impurity in diamond has an acceptor level
with binding energy 0.37 eV \cite{Fontaine}.
With increased doping, the boron impurity band will eventually overlap the 
diamond valence band and the system becomes metallic at a boron 
concentration of $8 \cdot 10^{20}$ cm$^{-3}$.
Since this is one order of magnitude lower than the doping at which 
superconductivity was observed \cite{Ekimov}, we felt justified in using a virtual
crystal approximation (VCA) in which the carbon nuclei have
charge $6-x$ and the crystal is neutral.

The valence bands were calculated with the scalar-relativistic
full-potential LMTO method \cite{Savrasov}, and the phonon dispersions and
the electron-phonon spectral function $\alpha^{2}F$ were
calculated with the linear-response method \cite{Savrasov}. Effects of
anharmonicity were considered in a second step. We used a triple-$\kappa$
$spd$ LMTO basis set and represented the charge densities and potentials by
spherical harmonics with $l\leq6$ inside non-overlapping muffin-tin spheres
and by plane waves with energies less than 400 Ry between the spheres. The
resulting band structure for undoped diamond agrees with those of
earlier LDA calculations. Due to the smallness of $k_{F},$ we needed to use a
fine $\mathbf{k}$-mesh chosen as a $1/32^{3}$ sublattice in reciprocal space.
$\alpha^{2}F$ is evaluated as a weighted sum over linewidths of individual phonons,
and for this a fine, yet affordable $\mathbf{q}$-mesh is needed. 
It was chosen as a $1/8^{3}$ sublattice in reciprocal space.
The $\lambda$-values obtained herewith are accurate when $x\gtrsim0.05,$ 
whereas Eq.~(\ref{e1}) which we have now derived analytically also in
three dimensions, is more accurate for smaller dopings.
Since, even for 10\% doping, we calculated an
increase of the lattice constant by less than a per cent, we did use
the experimental lattice parameters for the \emph{un}doped materials
in all subsequent calculations .

\begin{figure}[b]
\begin{center}
\includegraphics[width=0.9\linewidth,clip=true]{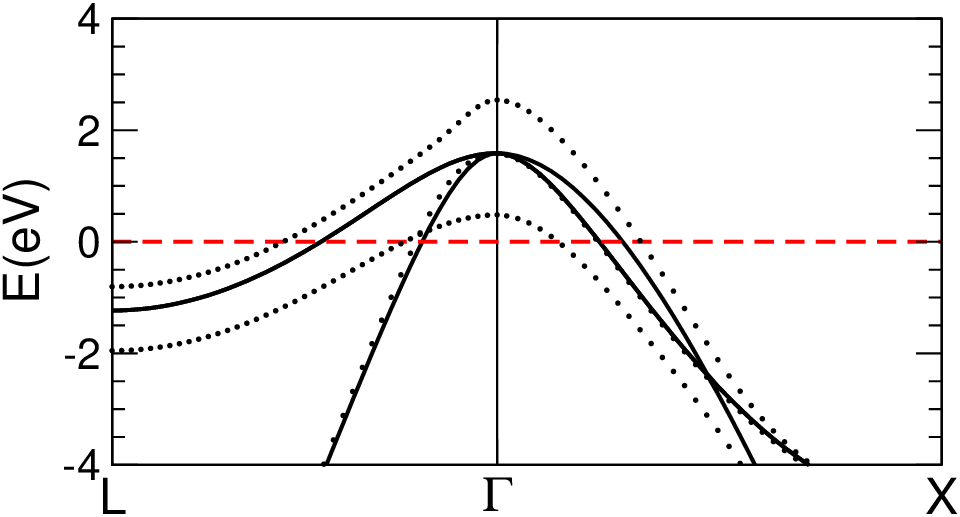}
\caption{\label{FP}
LDA band-structure of diamond with $x$=0.1 holes/C (full lines).  
A frozen optical zone-center phonon with 
two bonds stretched and two contracted by $u/\sqrt{2}$ splits
the bands by $\pm Du$ (dotted lines).
$u$=0.05{\AA}.}
\end{center}
\end{figure}

In Fig.~\ref{FP} we show the top of the valence-band structure 
calculated for 10\% hole-doped diamond. 
For this unrealistically heavy doping,  $N$ reaches 75\% of the 
$\sigma$-band DOS in MgB$_{2}.$ 
The electronic parameters may be found together with those for MgB$_{2}$ 
in the first columns of the table. 
Due to the deviations from parabolicity 
seen in Fig.~\ref{FP}, $N$
decreases somewhat faster than $x^{1/3}.$
As is well known, the LDA
gap is too small  and this leads to a slight
underestimate of the valence-band masses and the DOS. Nevertheless,
properties derived from the total energy, like phonon energies, are quite
accurate.

For the displacement $\mathbf{e}u$ of a frozen, optical zone-center phonon
with two bonds stretched and two contracted, say those in respectively the
positive and negative $z$-directions,
the top of the valence band is deformed as shown in Fig.~\ref{FP}: For small
$k$ there are two identical bands, split in energy by $\pm Du,$ and a band
which does not move with respect to the Fermi level. 
At $\Gamma$ the corresponding wavefunctions are those linear combinations
of the bond orbitals which have respectively
$(p_x \pm p_y)\sqrt{2}$ and $p_z$ symmetry. 
The value of
$D$ given in the table is seen to be nearly twice as large as in MgB$_{2}.$
For pure diamond it agrees with the accepted value \cite{defpot}, and it is
seen to decrease slightly with doping.

\begin{figure}[b]
\begin{center}
\includegraphics[width=0.9\linewidth,clip=true]{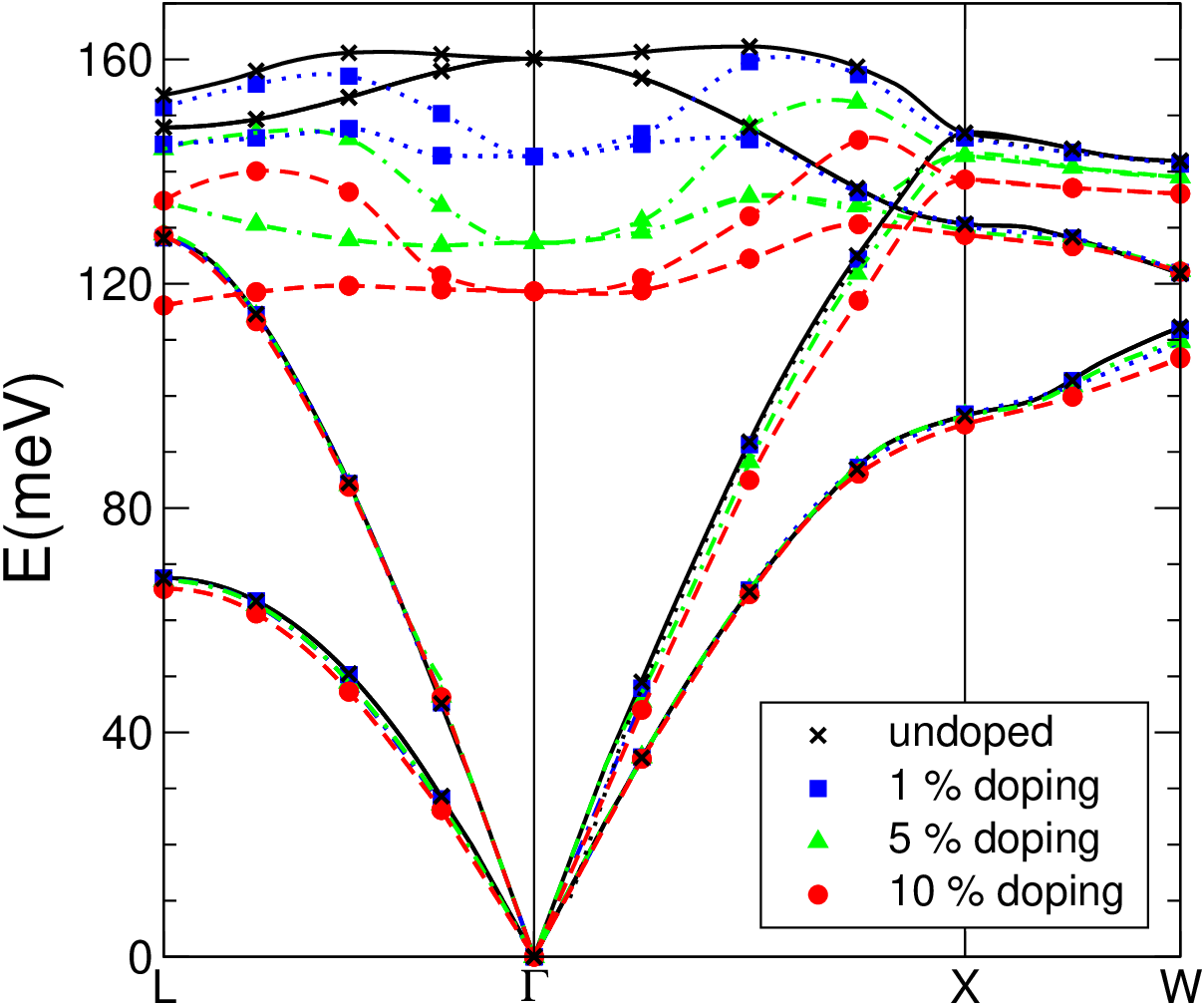}
\end{center}
\caption{Phonon dispersions calculated with the linear-response method for
diamond with $x=0,$ 1, 5, and 10 \% hole doping.}
\label{fig:phn}
\end{figure}

In Fig.~\ref{fig:phn} we show the phonon dispersions calculated in the
harmonic approximation for undoped and hole-doped diamond. The dispersions for
pure diamond, including the slight upturn of the uppermost mode when moving
away from the zone center \cite{phn:diam}, are well reproduced, and for the
frequency 1332 cm$^{-1}$ of the optical zone-center modes we calculate
$\omega_{0}$=1292 cm$^{-1}.$ Previous LDA calculations \cite{theo:diam} obtained
similar results. In the presence of hole-doping, the calculated dispersions of
the optical modes clearly exhibit softening near the zone center and a
3-dimensional Kohn anomaly around $q=2k_{F}.$

\begin{table*}[tp]
\caption{\label{tab}
$N$ is in states/eV/spin/f.u. $D$ is in eV/\AA . $\omega$ is in cm$^{-1}.$
$\lambda_{0}$ is the bare electron-phonon coupling constant defined in 
Eq.~(\protect\ref{e6}). 
$\lambda_{D}$ and $\lambda_{\omega}$ are estimates of the coupling constant 
as obtained from respectively Eq.~(\protect\ref{e1}) and the softening of $\omega$. 
$\lambda$ is obtained from the numerical linear-response calculation and includes 
all phonons and $\sigma$-electrons; for MgB$_{2}$ it is $\lambda_{\sigma\sigma}$ 
\protect\cite{GolubovJPCM}. $a\equiv E_{F}'/Dt$. 
$\lambda_{a}$ is $\lambda$ corrected for anharmonicity using Eq.(\ref{e3}). 
$T_{c}$ is obtained from Eq.~(\protect\ref{e4}) using $\lambda_{a},$ $\omega_{a},$ and 
$\mu^{\ast}\mathrm{=}0.1$.
}
\begin{ruledtabular}
\begin{tabular}{rcdcccccccd}
  & $N$  & D  & $\omega$ &
$\lambda_{0}$ &   $\lambda_{D}$ & $\lambda_{\omega}$ &
$\lambda$&  $a$ & $\lambda_a$ & T_c \\[1ex]
MgB$_2$ & 0.15& 12.4 &   536 & 0.33 & 1.01 & ---  & 1.02  &0.9 & 0.78 & 45       \\[1ex]
C  & 0.00 & 21.6 &  1292 & 0    &   0  & 0    &  0    &0.0 &  0   & 0       \\
3\%C  & 0.07 & 21.1 &  1077 & 0.21 & 0.30 & 0.33 & 0.30  &0.7 & 0.27 & 0.2     \\
5\%C & 0.08 & 20.8 &  1027 & 0.25 & 0.37 & 0.44 & 0.36  &0.9 & 0.33 & 2       \\
10\%C   & 0.11 & 20.4 &   957 & 0.32 & 0.57 & 0.62 & 0.56  &1.3 & 0.54 & 25       \\[1ex]
Si  & 0.00 &  6.8 &   510 &  0   &  0   & 0    &  0    &0.0 &  0   & 0       \\
5\%Si & 0.17 &  6.3 &   453 & 0.13 & 0.16 & 0.20 & 0.30  &1.4 & 0.30 & 0.3      \\
10\%Si  & 0.24 &  6.1 &   438 & 0.17 & 0.22 & 0.27 & 0.40  &2.0 & 0.40 & 3       \\[1ex]
Ge    & 0.00 &  5.8 &   317 & 0    &  0   &  0   &  0    &0.0 &  0   & 0       \\
10\%Ge    & 0.20 &  4.4 &   282 & 0.08 & 0.09 & 0.20 & 0.32  &5.1 & 0.32 & 0.4     \\
\end{tabular}
\end{ruledtabular}
\end{table*}

The softening of the zone-center phonons is $\left(  2/3\right)
\lambda$ instead of $\lambda,$ as in the case of MgB$_{2}.$ This is most
easily seen by considering a frozen phonon calculation and Fig.~\ref{FP}:
The adiabatic redistribution of $\left(  1/3\right)  2NDu$ electrons from the
upper third to the lower third of the deformed valence band decreases the
energy of each electron by $Du,$ and therefore perturbs the potential energy
of the harmonic oscillator by $-\left(  1/3\right)  2ND^{2}u^{2}$. As a
consequence, $\left(  1/2\right)  M\omega^{2}=\left(  1/2\right)  M\omega
_{0}^{2}-\left(  1/3\right)  2ND^{2},$ and by use of Eq.\ (\ref{e1}) we 
get: $\omega^{2}=\omega_{0}^{2} /  (1+2\frac{2}{3}\lambda)$. 
In MgB$_{2}$ no part of the $\sigma$-band is passive in the
screening of the phonon, so the factor 2/3 is missing.
The value of $\lambda$ deduced from the frequencies, 
$\omega$ and $\omega_{0}$ $\equiv$ $\omega (x\mathrm{=}0)$, 
of the optical zone-center modes calculated by linear response
is given in the table $(\lambda_{\omega})$.  This $\lambda_{\omega}$-value 
is seen to agree well with the value $\lambda_{D}$
obtained by use of Eq.~(\ref{e1}). In order to separate the materials
and dimensional dependencies of $\lambda,$ we express it in terms of a
\emph{bare} coupling constant, $\lambda_{0},$ and an enhancement due to the
phonon softening:
\begin{equation}
\lambda_{0}\equiv\frac{ND^{2}}{M\omega_{0}^{2}},\quad\mathrm{\lambda=}
\frac{\lambda_{0}}{1-2\frac{2}{3}\lambda_{0}},\quad\quad
\frac{\omega^{2}}{\omega_{0}^{2}}=1-2\frac{2}{3}\lambda_{0}.\label{e6}
\end{equation}
The enhancement is weaker in 3 dimensions than in 2, where the reduction
factor 2/3 is missing. As for the materials dependence, the 
$\lambda_{0}$-values given in the table first of all show that 
10\% doped diamond has the same $\lambda_{0}$ $\sim$ 1/3 as 
MgB$_{2}$: The bare force constant, $M\omega_{0}^{2}$, is 0.49 times 
its value in MgB$_{2}$, $N$ is 0.75, and $D$ is 1.65.
Due to the difference in dimensionality, $\lambda\sim\frac{1/3}{1-4/9}=0.6$ in
doped diamond, but $\lambda\sim\frac{1/3}{1-2/3}=1$ in MgB$_{2}.$ With
decreasing doping in diamond, $N$ decreases roughly like $x^{1/3},$ $D$
increases slightly, and $M\omega_{0}^{2}$ is constant. 
As a consequence, for 3\% doping $\lambda_{0}$ is only 0.21 
and $\lambda$ is 0.30.

\begin{figure}[t]
\begin{center}
\includegraphics[width=\linewidth,clip=true]{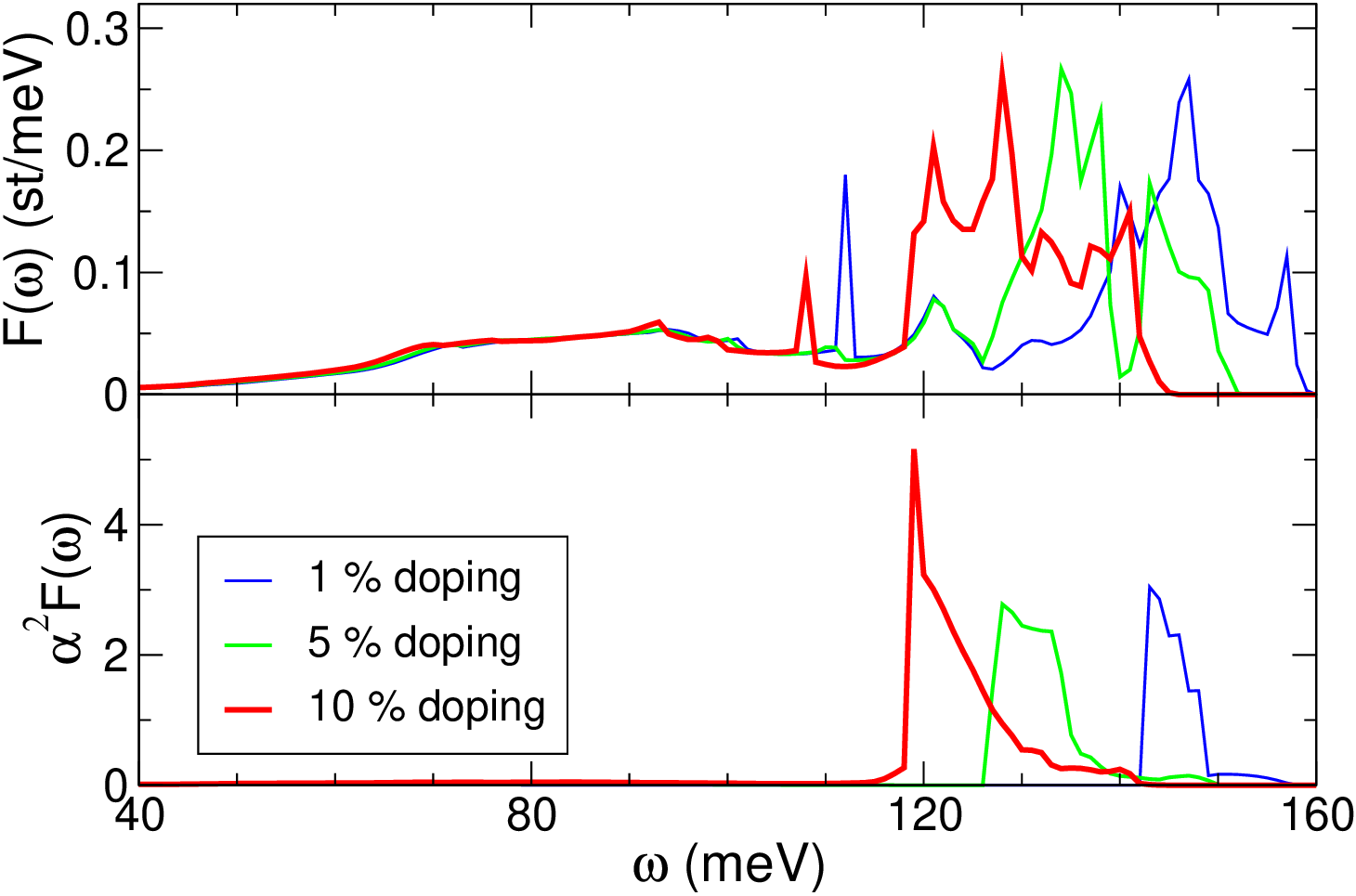}
\end{center}
\caption{The phonon density of states, $F(\omega),$ and Eliashberg function,
$\alpha^{2}F(\omega),$ calculated numerically by linear response considering 
all electrons and phonons.}
\label{fig:eli}
\end{figure}

We can also calculate the electron-phonon spectral function
and $\lambda=2\int\omega^{-1}\alpha^{2}F\left(
\omega\right)  d\omega$ \emph{numerically} by sampling over all 
phonon branches and energy bands. The result  shown in Fig.~\ref{fig:eli}
confirms that only the optical phonons interact with the holes: $\alpha^{2}F$
vanishes for phonon frequencies below that of the optical zone-center modes,
then jumps to a maximum, and finally falls. The decay occurs more slowly than
in MgB$_{2}$ due to the increase of dimensionality.  
The $\lambda$-values $\left(  \lambda\right)$ obtained from this calculation 
again agree well with those obtained from Eq.~(\ref{e1}) 
and from the phonon softening.

In MgB$_{2}$ the role of anharmonicity of the optical phonon modes with
$q$ $<$ $2k_{F}$ has been stressed\ \cite{Choi}; it hardens the phonon by about 20\%
and thus decreases $\lambda$ from 1.0 to $\lambda_{a}$=0.78, as given in the
table. However, this has recently been questioned \cite{Mauri}. 
While anharmonicity may be crucial in MgB$_2$ it has
at most a noticeable effect on the superconductivity in diamond at small
dopings ($x$ $<$3\%) as we shall now see:
The anharmonicity appears in frozen phonon calculations (see Fig.~\ref{FP}), 
because once the displacement $u$ exceeds $E_{F}/D,$ the lower band
is full so that the screening is lost \cite{Boeri}. In the expression for the
perturbation of the potential energy of the oscillator, $u^{2}$ should now be
substituted by $\left(  \left\vert u\right\vert -E_{F}/D\right)  ^{2}
\theta\left(  E_{F}/D-\left\vert u\right\vert \right)$, provided that we
simplify the DOS shape by a square.
It has been shown that the most important
anharmonic contribution to $T_{c}$ is the decrease of the first excitation
energy \cite{Hui}. 
By first-order perturbation theory, this is simply 
$\left(  1/3\right)  2ND^{2}t^{2}\left[1-\operatorname{erf}\left( 
 E_{F}/Dt\right)  \right]  ,$ 
where $t\equiv\sqrt{\hbar/M\omega}$
is the classical turning point in the ground state. 
Introducing again Eq.(\ref{e1}) we obtain the result:
\begin{equation}
\frac{\lambda_{a}}{\lambda}=\frac{\omega^{2}}{\omega_{a}^{2}}=\frac
{1}{1+2\left(  2/3\right)  \lambda\left[  1-\operatorname{erf}\left(
E_{F}/Dt\right)  \right]  }.\label{e3}
\end{equation}
For MgB$_{2}$ the assumption of a square $N\left(  E\right)  $ is good, but
due to the missing factor 2/3 and the presence of the $\pi$-band, 
$\left( 2/3\right)  $ in Eq. (\ref{e3}) should be substituted by 
$[1-N_{\pi}/\left(N+2N_{\pi}\right)]$. For hole-doped diamond, $N(E)$ has
square-root shape, and this we crudely take into account by substituting
$E_{F}$ in Eq. (\ref{e3}) by $E_{F}'=\left(  2/3\right)  E_{F}$. 
In the table we have included the ratio $E_{F}'/Dt\equiv a$ as well 
as the results  for $\lambda_{a}$.
We see that the effect of anharmonicity may be important in MgB$_{2}$ but
merely noticeable in hole-doped diamond \cite{anharm:semi}.

For this type of superconductivity, which is characterized by an Eliashberg
function with the shape exhibited in Fig.~\ref{fig:eli}, and which we can
idealize by a $\delta$-function at the frequency $\omega$ of the optical
zone-boundary phonon, solution of the Eliashberg equations yields with high
accuracy:
\begin{equation}
T_{c}=\omega\exp\left(  \frac{-1}{\frac{\lambda}{1+\lambda}-\mu^{\ast}
}\right)  .\label{e4}
\end{equation}
This is McMillan's expression with all numerical factors, which he obtained by
fitting to $F\left(  \omega\right)  $ of niobium, set equal to unity. For the
cases considered in the present paper, it does not make much difference
whether one uses McMillan's factors or unity inside the exponential, but it is
important that the prefactor is $\omega,$ rather than $\left\langle
\omega_{\ln}\right\rangle /1.2.$

We can finally estimate $T_{c}$ from
 Eq.~(\ref{e4}) with the values for  $\lambda_{a}$ given in the table
and $\omega_a$ from Eq.~(\ref{e3}).
For the Coulomb pseudopotential, the standard value 
$\mu^{\ast}$=0.1 was used in all cases. 
For MgB$_{2}$ we neglected the $\pi$-bands.
Considering the uncertainties in our calculation of $\lambda$ and $\omega,$
the uncertainty of $\mu^{\ast},$ and the experimental estimation of the doping
level, we do find critical temperatures in good agreement with present
experimental knowledge. We therefore believe to have substantiated our claim
that the superconductivity in hole-doped diamond is of MgB$_{2}$-type, but in
three dimensions.

We repeated our calculations for hole-doped Si and Ge, and include those
results in the table for which $E_{F}$ largely exceeds the spin-orbit
splitting, which we neglected. Whereas hole-doped C shows superconductivity
above 1 K for doping levels presently obtainable, Si and Ge seem 
to need twice as high doping levels. 
The main reason is that the deformation potential in Si
and Ge is about four times smaller than in C,  
which is too small to take advantage of having twice as
large a DOS and a three times smaller force-constant.
There is also a qualitative difference to diamond: For
heavily doped Si and Ge, the holes not only couple to the optical, but also to
the acoustic phonons. This is the reason why $\lambda$  
exceeds $\lambda_{D}\sim\lambda_{\omega}$. 

In conclusion, we have shown that the recently discovered superconductivity in
hole-doped diamond below 4K is of MgB$_{2}$-type, but in three dimensions.
This means that the mechanism is coupling of a few holes at top of the
$\sigma$-bonding valence band to the optical bond-stretching 
zone-center phonons. The increase
from 2 to 3 dimensions limits the strong softening of the optical modes mainly
responsible for the high $T_{c}$ in MgB$_{2}.$ On the other hand, the
deformation potentials in diamond are twice stronger than in MgB$_{2}$. 
Kelvin-range superconductivity in Si and Ge would require
hole-doping levels of 5-10\%. Finally, we have obtained simple analytical
expressions for MgB$_{2}$-type superconductivity.

A purely electronic mechanism for the observed superconductivity
was recently suggested \cite{Baskaran}, and 
after submission of the present manuscript two works similar to
ours appeared  \cite{Pick:diam,Xiang}.
The latter used a supercell approach to simulate the boron doping
and found an electron-phonon coupling 
in very good agreement with our results.

We are grateful to O. Dolgov, 
M. Cardona, G. B. Bachelet, E. Cappelluti, and L. Pietronero for many 
interesting discussions.

\end{document}